\documentclass[10pt]{article}

\usepackage{authblk}
\usepackage[square,numbers]{natbib}
\usepackage[T1]{fontenc}
\usepackage[utf8]{inputenc}
\usepackage{amsmath,amssymb,amsfonts}
\usepackage{algorithmic}
\usepackage{graphicx}
\usepackage{subcaption}
\usepackage{textcomp}
\usepackage{url}
\usepackage{xurl}
\usepackage[htt]{hyphenat}
\usepackage{xcolor}
\usepackage{array}
\usepackage{booktabs}
\usepackage{braket}
\usepackage{enumitem}
\usepackage{ifthen}
\usepackage{makecell}
\usepackage{multirow}
\usepackage{pifont}
\usepackage{ragged2e}
\usepackage{tcolorbox}
\usepackage{xspace}
\usepackage[hidelinks]{hyperref}
\usepackage{geometry}

\def\BibTeX{{\rm B\kern-.05em{\sc i\kern-.025em b}\kern-.08em
    T\kern-.1667em\lower.7ex\hbox{E}\kern-.125emX}}

\definecolor{light-gray}{gray}{0.80}

\newcommand{\Description}[1]{}

\newcolumntype{C}[1]{>{\centering\arraybackslash}p{#1}}

\newcommand{\ApproachName}{\textsc{QPipe}\xspace}

\newcommand{\TCO}{TCO\xspace}
\newcommand{\parseAgent}{\textsc{Parse Agent}\xspace}
\newcommand{\analysisAgent}{\textsc{Analysis Agent}\xspace}
\newcommand{\blueprintAgent}{\textsc{Blueprint Agent}\xspace}
\newcommand{\encodingAgent}{\textsc{Encoding Agent}\xspace}
\newcommand{\codegenAgent}{\textsc{Codegen Agent}\xspace}
\newcommand{\reviewerAgent}{\textsc{Review Agent}\xspace}
\newcommand{\executionSandbox}{\textsc{Execution Sandbox}\xspace}
\newcommand{\combinationAgent}{\textsc{Combination Agent/Script}\xspace}
\newcommand{\verificationAgent}{\textsc{Verification Agent}\xspace}
\newcommand{\AblationSingle}{\textsc{single}\xspace}
\newcommand{\AblationWithoutSkill}{\textsc{w/o skill}\xspace}
\newcommand{\AblationWithoutKnowledge}{\textsc{w/o knowledge}\xspace}
\newcommand{\AblationWithoutReview}{\textsc{w/o review}\xspace}
\newcommand{\AnswerToRQ}[2]{%
  \begin{tcolorbox}[
    colback=black!1!white,
    colframe=black!40!white,
    left=0.25mm,
    right=0.25mm,
    top=0.25mm,
    bottom=0.25mm,
    boxsep=0.66mm,
    arc=0.1mm,
  ]
  \textbf{Answer to RQ#1:} #2
  \end{tcolorbox}
}

\newboolean{showcomments}
\setboolean{showcomments}{true} 

\ifthenelse{\boolean{showcomments}}{
	\newcommand{\nbc}[3]{%
		{\colorbox{#3}{\bfseries\sffamily\scriptsize\textcolor{white}{#1}}}%
		{\textcolor{#3}{\sf\small$\langle$\textit{#2}$\rangle$}}%
	}
	
}{
	\newcommand{\nbc}[3]{}

}

\title{QPipe: LLM-Based Agentic Quantum Application Generation for Test Optimization}
\author[1,3]{Ming Tao}
\author[1,*]{Yuechen Li}
\author[1,*]{Tao Yue}
\author[1]{Man Zhang}
\author[2]{Aitor Arrieta Marcos}
\affil[1]{Beihang University\\
\texttt{\{taoming, liyuechen, yuetao, manzhang\}@buaa.edu.cn}}
\affil[2]{Mondragon University\\
\texttt{aarrieta@mondragon.edu}}
\affil[3]{QuantumJump.ai}
\affil[*]{Corresponding author}
\date{}

\begin{document}

\maketitle

\begin{abstract}
Quantum computing is increasingly explored for software engineering (SE) optimization, but translating natural-language (NL) task-level requirements into executable quantum applications still demands substantial quantum and programming expertise. We present \ApproachName, a large language model (LLM)-based multi-agent architecture that autonomously turns NL requirements into traceable quantum-application workflows through specialized agents for requirement parsing, formulation, code generation, review, execution, and verification.
We evaluate \ApproachName on 20 NL requirements, each associated with a real-world benchmark and a test-optimization problem.
\ApproachName successfully completes the key stages of quantum-application generation across requirements, achieving average rates of 100\% for code compilation and 96.7\% for application execution and final-result combination, with average generation costs of 260.1 seconds and 1.89M tokens per requirement. Among the generated quantum applications that execute successfully, the returned solutions outperform the offline genetic algorithm baseline in most cases.
Ablation results further show that \ApproachName's advantage depends on retaining code-generation skills, task knowledge, review feedback, and multi-agent decomposition.
These results indicate that agentic coordination can 
support generation of executable quantum applications
for tackling test optimization problems from real-world benchmarks.
\end{abstract}


\section{Introduction}\label{sec: intro}
Quantum computing (QC) is increasingly explored for software engineering (SE) optimization problems~\cite{zhang2026quantum}, many of which are classically studied in search-based software engineering (SBSE)~\cite{harman2012search}. Recent work has applied quantum optimization to SE tasks such as \TCO, including test case minimization (TCM) with quantum annealing (QA), which searchs low-energy states of an encoded objective, and test case selection (TCS) with quantum approximate optimization algorithm (QAOA)~\cite{wang2024test,wang2024quantum}. These results suggest that quantum optimization is becoming a plausible execution target for SE tasks, but still requires substantial expertise to use effectively.

The remaining challenge is to translate an NL requirement for an SE optimization task into an executable quantum application, including problem formulation, encoding, solver selection, orchestration, and execution. Existing works on quantum workflow engineering offer useful abstractions, including workflow modeling in \textit{QuantME}~\cite{weder2020quantme}, pattern-based workflow construction~\cite{beisel2025pattern}, and feasibility-oriented design in \textit{Q-READY}~\cite{yue2026qready}. In parallel, large language models (LLMs) have shown promise in benchmarked quantum code generation~\cite{vishwakarma2024qiskit,guo2025quanbench,basit2025qhackbench}, QAOA circuit generation~\cite{tyagin2025qaoa}, and NL-to-quadratic unconstrained binary optimization (QUBO) transformation~\cite{zhang2025llm}. However, these efforts do not directly address the complete requirement-to-application path for SE optimization tasks: they do not take both the task definition and its concrete instance as input and return a runnable application.

To address this gap, this paper presents \ApproachName, an LLM-based agentic system that takes an NL specification of an SE optimization problem and its concrete instance data and generates a runnable hybrid quantum application with an application-level result. The generated artifact therefore includes the problem-to-program path that is absent from isolated circuit or code-fragment generation.
In fact, generating such an application asks the agentic system to answer four questions specific to quantum computing: 
(1) \textit{Can the optimization problem, or a part of it, be addressed by a suitable quantum algorithm?} This determines whether the task has a quantum-suitable component to execute. 
(2) \textit{How should the problem and instance data be encoded for that algorithm?} The encoding must represent the task's objectives and constraints in a form the selected algorithm can process. 
(3) \textit{Which functions should run in classical subroutines, and which optimization steps should run in quantum subroutines?} Unlike a pure classical application, a hybrid quantum application should define how classical processing connects to quantum execution. 
(4) \textit{Can the expected quantum execution cost fit within the available resource budget?} Even a valid formulation may be infeasible if its qubit, runtime, or simulation demands exceed the target environment's limits. 

The resulting workflow uses distinct stages whose artifacts carry these decisions forward. The \analysisAgent identifies an optimization kernel that can be handled by a suitable quantum algorithm, specifies its subproblem formulation and classical data interface, and separates it from functions that remain classical (Question (1)). The \encodingAgent then represents the task objectives, constraints, and instance data in the formulation required by the selected algorithm (Question (2)). The \blueprintAgent organizes the kernel and its supporting classical functions into a hybrid workflow. These functions may include data preparation, constraint handling, postprocessing, and result interpretation; the blueprint specifies how they interact with quantum execution (Question (3)). The \blueprintAgent and \encodingAgent also assess feasibility under the target resource budget: when an original instance exceeds the qubit limit, the encoding stage decomposes it into executable subproblems and records how their outputs should be combined (Question (4)). The \codegenAgent realizes this plan as a program, the \reviewerAgent checks and repairs its semantic consistency, and the execution, combination, and verification stages run the subproblems, produce the application-level result, and record supporting evidence.
Also, \ApproachName supplies curated task knowledge and coding skill to enhance the workflow effectiveness, where task knowledge helps the workflow ground the NL requirement in quantum formulations of underlying quantum algorithms, while coding skill guides the implementation of those formulations as complete programs. 
Intermediate artifacts make the path from the NL task and instance to the generated application inspectable. 

We instantiate and evaluate \ApproachName on \TCO through two problem variants, TCS and TCM, using 10 benchmark instances from prior quantum-optimization studies~\cite{wang2024quantum,wang2024test,yang2026ising} and 20 NL requirements, each associated with one benchmark instance and one of the two variants.
This evaluation measures generation cost and effectiveness, analyzes workflow patterns, compares solution quality against a genetic algorithm (GA) baseline, assesses ablations that remove code-generation skills, task knowledge, review feedback, and multi-agent decomposition, and compares three backbone LLMs.
Results show that \ApproachName completes key quantum-application generation stages: it compiles applications for all requirements and completes execution and final combination for 96.7\% of them, with average costs of 260.1 seconds and 1.89M tokens per requirement. Executable applications generated by \ApproachName produce solutions that exceed the GA baseline in most cases. Workflow traces show that \ApproachName's agents autonomously converge on QUBO+QAOA formulations and script-based final aggregation, while varying decomposition and orchestration structure across runs. Ablations show that code-generation skills, task knowledge, review feedback, and multi-agent decomposition all contribute to performance, with the largest drop from removing code-generation skills. In the backbone comparison, Claude performs strongest, DeepSeek remains competitive in early stages and compilation, and Llama fails earlier at code generation. Broader empirical studies are needed to generalize these model-level observations.


The main contributions of this paper include:
\begin{itemize}[leftmargin=*,topsep=0pt]
    \item We present \ApproachName as a preliminary step toward LLM-based agentic generation of quantum applications: the system converts an NL SE optimization requirement and its concrete instance into a runnable hybrid quantum application with a traceable generation record.
    \item We integrate quantum-specific steps, task knowledge, and coding skill into \ApproachName to facilitate the generation of executable and useful quantum applications under the practical constraints.
    \item We evaluate \ApproachName on 20 NL requirements and 3 LLMs, exhibiting~\ApproachName's desirable performance in terms of execution cost and effectiveness, solution quality, and component contribution.
    \item We publicly release a reusable artifact of \ApproachName at~\url{https://github.com/marvintao/QSE_workflow_artifact} and the empirical-study data at~\url{https://github.com/marvintao/QSE_workflow_empirical_study} to support future research and enhance reproducibility.
\end{itemize}

The remaining of this paper is organized as follows. Section~\ref{sec: background} introduces the background; Section~\ref{sec: methodology} presents \ApproachName; Section~\ref{sec: empirical} discusses its evaluation; Section~\ref{sec: related} presents related work; and Section~\ref{sec: conclusion} concludes the paper.

\section{Background}\label{sec: background}

Quantum applications often combine quantum computation with classical
orchestration. 
Unlike classical programs, quantum programs operate on quantum states and return probabilistic classical outcomes after measurement.
For example, a qubit state can be written as
$|\psi\rangle=\alpha|0\rangle+\beta|1\rangle$ with
$|\alpha|^2+|\beta|^2=1$, and measuring in the computational basis returns $0$ with probability $|\alpha|^2$ and $1$ with probability $|\beta|^2$.
Thus, classical orchestration is needed to prepare inputs, configure execution, collect samples, and interpret results.
In the noisy intermediate-scale quantum (NISQ) era,
limited qubit counts, noise, backend constraints, and access cost further make such orchestration central to practical quantum applications.
Hybrid workflows can take different forms; in one common pattern, classical code encodes an input instance into a quantum-compatible representation, configures the solver or optimizer, executes it on a quantum simulator or hardware backend, and post-processes sampled or optimized outputs.

Optimization is a common target for quantum applications in the NISQ era.
A discrete optimization problem is often encoded as a QUBO objective or an
equivalent Ising energy:
\begin{equation}
    \min_{x\in\{0,1\}^n} x^\top Qx
    \quad\Leftrightarrow\quad
    \min_{s\in\{-1,+1\}^n}\sum_i h_i s_i+\sum_{i<j}J_{ij}s_i s_j .
    \label{eq:qubo_ising}
\end{equation}
In Eq.~\eqref{eq:qubo_ising}, $x_i$ is a binary decision variable and $Q$
stores linear and pairwise costs or penalties; equivalently, $s_i$ is a spin
variable, $h_i$ gives its local bias, and $J_{ij}$ gives the interaction between
two decisions.
In SE optimization, these variables may represent decisions such as selecting a test case, while coefficients encode costs, benefits, interactions, and constraint-violation penalties. QAOA realizes this objective with a parameterized gate-based circuit optimized by a classical optimizer.
QA maps the same style of objective to an energy landscape and searches for low-energy states. These routines usually produce candidate assignments or samples, not guaranteed optima. Recent work has applied such formulations to SE optimization problems~\cite{zhang2026quantum}, such as \TCO~\cite{wang2024quantum}.



SBSE formulates SE tasks as optimization problems over discrete choices, objectives, and constraints~\cite{harman2012search}.
We study \TCO because it is well established in both classical SBSE and quantum optimization, and provides formulations and benchmarks for evaluation~\cite{yoo2012regression, hu2024test, wang2024quantum, wang2024test, yang2026ising, zhang2026quantum}.
Below, we present two \TCO problems defined in~\cite{wang2024quantum}:
\begin{itemize}[leftmargin=*]
    \item TCM: Given a test suite $T = \{t_0, \cdots, t_{n-1}\}$ with testing attributes and objectives $F = \{f_0, \cdots, f_{m-1}\}$, TCM selects a minimum subset $T' \subseteq T$ that satisfies all objectives as much as possible. These objectives include attribute-related objectives and a specific objective to minimize $|T'|$.
    \item TCS: Given a test suite $T = \{t_0, \cdots, t_{n-1}\}$ with testing attributes and objectives $F = \{f_0, \cdots, f_{m-1}\}$, TCS selects a subset $T' \subseteq T$ that maximizes objective satisfaction. Each objective is linked with a test case attribute.
\end{itemize}

\section{Approach}\label{sec: methodology}

\begin{figure}[!t]
    \centering
    \includegraphics[width=0.7\columnwidth]{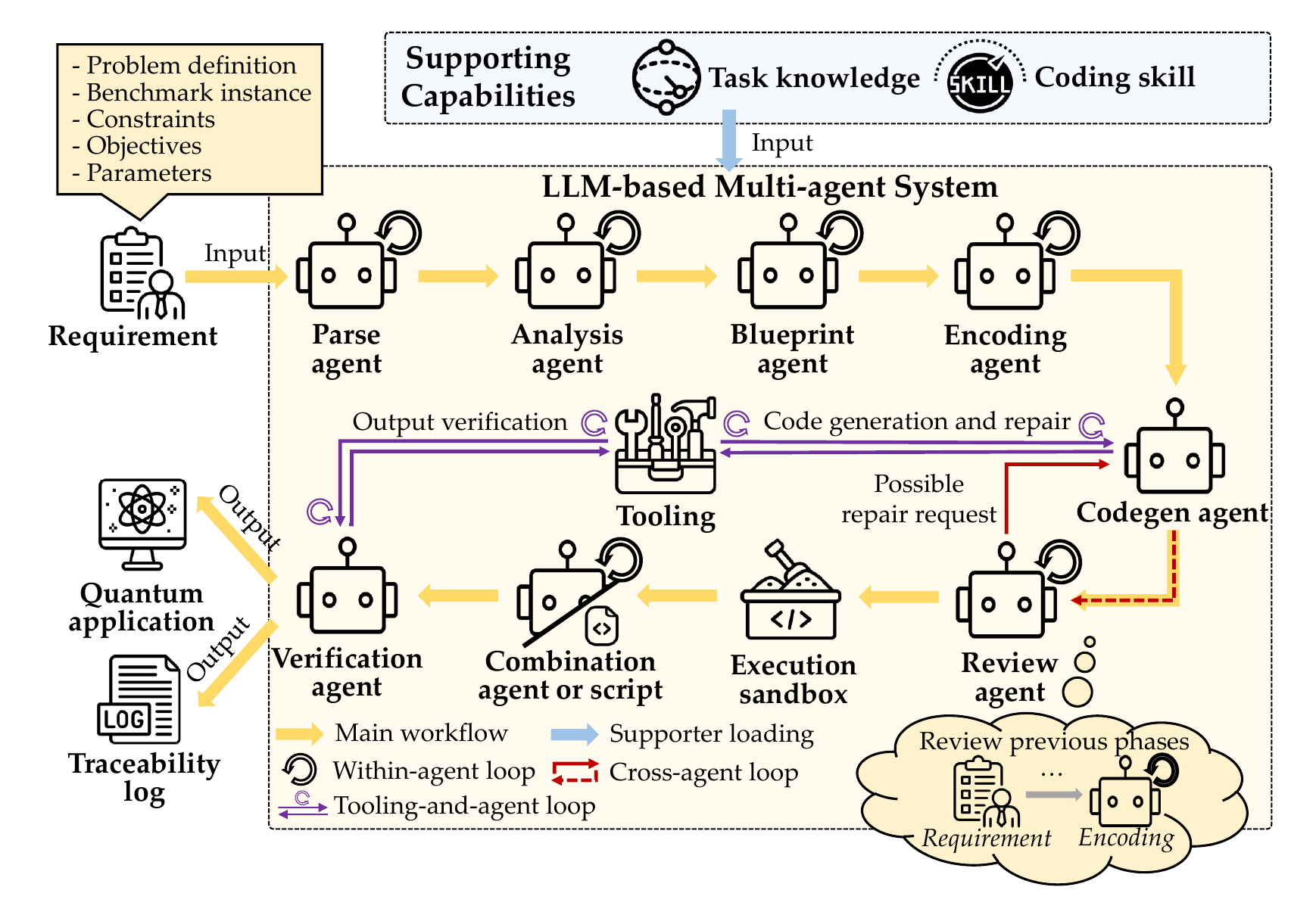}
    \caption{Overview of \ApproachName.}\label{fig:workflow}
\end{figure}

An overview of \ApproachName is given in Fig.~\ref{fig:workflow}, from where we can see that \ApproachName takes an NL requirement as input and produces two outputs: an executable quantum application and a generation record (Section~\ref{subsec:inputoutput}).
%
\ApproachName uses 8 specialized agents to manage the transition from an NL requirement to executable code.
\parseAgent extracts the SE task, objectives, constraints, and project data from the requirement.
\analysisAgent assesses whether and how the task can be formulated as a quantum-suitable optimization problem, e.g., a QUBO or Ising model.
\blueprintAgent turns this analysis into an application-level plan, specifying the formulation, workflow structure, solver choice, and expected artifacts.
\encodingAgent constructs the concrete optimization encoding required by the selected quantum formulation.
\codegenAgent generates candidate implementations from the blueprint and encoding.
Next, \reviewerAgent inspects the artifacts from previous agents and requests repairs when it detects inconsistencies, execution risks, etc.
After review passes, the generated quantum application is executed in \executionSandbox, and \combinationAgent combines decomposed subproblem outputs when needed. In the end, \verificationAgent verifies the result against both the quantum application and the generation record.

This architecture of \ApproachName is motivated by the heterogeneous nature of requirement-to-application generation. Translating a NL requirement into an executable quantum application requires requirement interpretation, quantum-suitability analysis, workflow planning, encoding, code generation, and consistency checking. Assigning these steps to specialized agents makes the process more controllable, enables targeted feedback and repair, and preserves traceability across intermediate artifacts. Moreover, this modular design also supports independent improvement of individual agents, since their prompts, skills, tools, models, and verification strategies can be updated separately. The detailed discussions of these agents are provided in Section~\ref{subsec:agents}.
Agents invoke tools to obtain executable feedback during generation, execution, and verification (Section~\ref{subsec:tooling}), while
%
%
supporting capabilities provide curated task knowledge and coding skill used throughout the workflow (Section~\ref{subsec:supporting_capabilities}).

\subsection{Input and Output of \ApproachName}
\label{subsec:inputoutput}
\ApproachName takes an NL requirement for an SE optimization task as input, and returns an executable quantum application together with a generation record, as discussed below.

\subsubsection{Input: Task-level Requirements}\label{sec: user_requirements}
The problem definition names the optimization task such as \TCO.
\ApproachName takes task-level requirements in NL as input. 
Each requirement specifies an externally posed SE optimization task for the generated application, where the requirement template includes the target problem, benchmark instance, presence of constraints, optimization objectives, and potential predefined parameters. Note that these requirements are not full industrial requirements specifications that capture stakeholder elicitation or complete system-level constraints; rather, they define the intended functionality, problem scope, and input conditions of the quantum application to be developed.
The benchmark instance grounds the task in concrete project data, e.g., a test suite with per-case attribute fields (e.g., execution cost).
Constraints delimit the feasible search space: they include minimum coverage for the TCS variant and target subset size for the TCM variant, as well as non-negativity of selection indicators.
Objectives specify the search targets to minimize, maximize, or balance, such as execution cost or required coverage.
Parameters provide task-specific values stated in the requirement, such as requested solution sizes, regularization coefficients, or per-objective weights; in our \TCO instances, these weights are non-negative scalars attached to each objective, such as cost weight and diversity weight, that combine multiple criteria into a single scalarized objective.
For example, a TCS requirement may ask for a subset of tests from a given test suite, provide fields such as execution cost and fault-detection rate, specify a required coverage threshold, and assign weights to the optimization objectives.

Together, these elements capture the task, search space, and optimization criteria in the same way SE optimization problems are typically specified in practice, while \ApproachName handles quantum formulation, encoding, and feasibility constraints.

\subsubsection{Output: Generated Quantum Application}
\ApproachName outputs an executable quantum application as a Python~3 program built on Qiskit and Qiskit-Aer in \executionSandbox. The artifact is a complete runnable program rather than an isolated circuit or code fragment: it loads the given instance, constructs the optimization model, invokes a quantum or hybrid routine, postprocesses the returned result, and emits a numeric output in a machine-readable form.
For example, a generated application for TCS may define a \texttt{main()} entry point that reads the inlined test-suite instance, builds a QUBO matrix, runs QAOA on the Qiskit-Aer simulator under the qubit budget, decodes the returned bitstring into a selected test subset, and prints the reported value and supporting result data. This application-level output remains inspectable and re-runnable, and the same output schema can accommodate alternative encodings, such as Ising or Hamiltonian forms, and different simulators or hardware backends.

\subsubsection{Output: Generation Record}
\label{subsubsec:generation_record}
\ApproachName also returns a generation record for each generated application.
Rather than exposing low-level storage fields, the generation record captures
the artifact produced at each stage so that the requirement-to-application
chain is reproducible: \parseAgent records the structured task description,
including objectives, constraints, instance schema, and ambiguity flags;
\analysisAgent records the kernel specification, including formulation type,
classical/quantum split, and qubit budget; \blueprintAgent records the workflow
blueprint for preparation, quantum execution, and postprocessing;
\encodingAgent records the per-subproblem encoding plan and combination rule;
\codegenAgent records the candidate program and the tool feedback used to
produce it; \reviewerAgent records its verdict and repair feedback;
\combinationAgent records per-subproblem outputs and the aggregated value; and
\verificationAgent records the final verdict, such as
\texttt{cross\_checked}, \texttt{unverified}, \texttt{mismatch},
\texttt{off\_spec}, or \texttt{failed}, together with supporting evidence.
This per-agent view mirrors the workflow and helps identify where a generated
application diverges from the requirement.

The generation record provides an audit trail from the submitted requirement to
the final output. Since an LLM-assisted run may fail during compilation,
execution, verification, or repair, the record preserves the artifacts,
outcomes, and evidence needed to analyze cost, executability, and failure
location. It serves as an evidence artifact rather than a correctness proof.
\subsection{Tooling of \ApproachName}
\label{subsec:tooling}
Tooling is important in \ApproachName because \codegenAgent needs concrete execution feedback to generate and repair runnable programs, while \verificationAgent needs evidence from generated artifacts and execution outputs to check whether the application implements the intended optimization formulation and whether its reported value can be independently validated.
As shown in Table~\ref{tab:tooling}, \ApproachName exposes 7 tools for code generation and 6 for verification. Agents are restricted to these locally provided tools, so the evaluation focuses on whether the workflow can reason over the supplied requirement, curated knowledge, and executable feedback, rather than retrieve arbitrary external material.

For code generation, the tools give \codegenAgent access to runtime knowledge,
candidate-program state, and executable feedback. They provide
\texttt{environment.md}, which records the Python, Qiskit, and Qiskit-Aer
versions and sandbox caps on qubits, time, and memory;
\texttt{qiskit\_cookbook.md}, which gives curated patterns for parameterized
circuits, transpilation, and measurement; and a building-blocks API manifest,
which documents reusable QAOA cost layers, CNF clause encoders, and Hamiltonian builders. They also store the current candidate program so repair updates one artifact rather than regenerating from scratch. Executable feedback comes from \texttt{static\_check} for unsafe
imports, file I/O, and subprocess use; \texttt{compile\_check} for syntax and
import errors; and \texttt{run\_in\_sandbox}, which executes the candidate on a
sample subproblem under the same qubit and time caps used in full execution.

For verification, the tools let \verificationAgent inspect artifacts, check
results against a classical reference, and record the final verdict, thereby
assessing formulation alignment, result consistency, and verification status.
Reference checking centers on \texttt{solve\_instance\_classically}, which
dispatches on the instance schema produced by the program: a
Hamiltonian instance triggers exact eigendecomposition of the
corresponding Pauli-sum operator to obtain a reference objective value, while a
QUBO instance with up to 20 binary variables triggers brute-force
enumeration of $\mathbf{x}^{\top}M\mathbf{x}$ over all $2^n$ assignments. If
the produced instance does not match a supported verification format, the tool
returns \texttt{unverified}, so the verdict reflects only available evidence.

\begin{table}[!t]
	\centering
	\scriptsize
	\renewcommand{\arraystretch}{0.8}
	\caption{Tools exposed to \codegenAgent and \verificationAgent.}
	\label{tab:tooling}
	\resizebox{0.98\columnwidth}{!}{

\begin{tabular}{@{}p{0.15\columnwidth}lp{0.52\columnwidth}@{}}
	\toprule
	Agent & Tool interface & Purpose \\
	\midrule
	\multirow{7}{=}{\codegenAgent}
	& \texttt{read\_environment\_facts} & Inspect Python/Qiskit and sandbox capabilities \\
	& \texttt{read\_qiskit\_cookbook} & Retrieve Qiskit construction patterns \\
	& \texttt{read\_blocks\_api} & Retrieve encoding/evaluation APIs \\
	& \texttt{write\_kernel\_code} & Store candidate program \\
	& \texttt{static\_check} & Reject unsafe imports/calls \\
	& \texttt{compile\_check} & Report syntax/import errors \\
	& \texttt{run\_in\_sandbox} & Execute sampled case and return feedback \\
	\midrule
	\multirow{6}{=}{\verificationAgent}
	& \texttt{read\_kernel\_spec} & Inspect formulation, interface, and qubits \\
	& \texttt{read\_code} & Inspect generated implementation \\
	& \texttt{read\_result\_instance} & Inspect solved instance \\
	& \texttt{read\_result\_value} & Inspect reported objective value \\
	& \texttt{solve\_instance\_classically} & Compute exact reference if feasible \\
	& \texttt{verdict} & Record final verification outcome \\
	\bottomrule
\end{tabular}
	}
\end{table}
\subsection{Agent Roles of \ApproachName}
\label{subsec:agents}

\ApproachName uses role separation to decompose requirement interpretation, quantum planning, code construction, and result checking into distinct agent responsibilities.
Each agent role is intentionally narrow: it sees only the artifacts it needs, emits a single named artifact type, and is constrained by its own prompt template, available tools, etc., and, where applicable, within-agent repair loop. The roles fall into three groups: \parseAgent, \analysisAgent, \blueprintAgent, and \encodingAgent consume upstream artifacts and emit one new intermediate artifact; \codegenAgent and \verificationAgent carry explicit tool loops; and \reviewerAgent implements a cross-agent repair loop.
This role separation is needed because SE optimization requires the workflow to preserve requirement semantics during decomposition, maintain objective alignment during code generation, and keep combined results consistent with the full problem. The remainder of this section introduces the eight roles in workflow order.

\subsubsection{Requirement Interpretation and Quantum Encoding}\label{sec: requirements and encoding}

\textbf{\parseAgent} records what is stated in the requirement, such as the optimization objective and available instance data, and flags ambiguity.
Because its role is limited to problem interpretation, it structures the requirement without making solution-side decisions, such as selecting a quantum algorithm or inventing missing constraints.

\textbf{\analysisAgent} performs the quantum-specific problem decomposition. It
identifies the quantum kernel, the optimization core to be executed on
the quantum node, and states the subproblem formulation handled by that kernel,
its classical data interface, and the feasible qubit budget. This role turns
an NL requirement into a quantum optimization task by deciding which
part remains classical and which part is mapped to the quantum node. The
decomposition is necessary as quantum execution is not a direct
replacement for the whole program; only the part that can be encoded as a
quantum objective should be sent to the quantum node, while data handling,
constraint preparation, and result interpretation remain classical.

\textbf{\blueprintAgent} turns the selected kernel into a hybrid workflow plan.
It organizes classical preparation, the quantum node, and postprocessing as a
workflow graph, with data-flow edges and a feasibility estimate. The generated artifact is therefore not merely a circuit, but a plan for a hybrid program in
which project data are prepared classically, the quantum kernel is executed
under resource limits, and the returned value is interpreted by classical
postprocessing.

\textbf{\encodingAgent} is the main task-specific role in this group. Because
quantum execution is constrained by qubit budgets, optimization instances
often cannot be passed to the selected kernel at full scale. Rather than
sampling a smaller problem, \encodingAgent maps the full instance to
executable subproblems. It produces one subproblem when the instance fits the
budget, or decomposes decision variables or instance rows into nonoverlapping
blocks when it does not. It also records the combination rule that later
stages must follow. This makes explicit how the full requirement is mapped to
smaller quantum executions under resource limits while preserving SE-level
traceability.


Take the \texttt{elevator\_o2} benchmark as an example (details shown in Table~\ref{tab: benchmark}). 
For the TCS requirement, \parseAgent extracts the objectives from the requirement: test-execution cost $F_{\mathrm{cost}}$ and passenger-traffic diversity $F_{\mathrm{div}}$; \analysisAgent selects a QUBO/QAOA kernel; and \encodingAgent maps the selection of each test case $t_i$ to a binary variable $x_i \in \{0,1\}$, resulting in $\min_x \sum_i (F_{\mathrm{cost}}(t_i)-F_{\mathrm{div}}(t_i))x_i$.
As for the corresponding TCM requirement, \parseAgent also identifies subset size, $F_{\mathrm{size}}(T')=|T'|$, as an objective; differently, \encodingAgent adds a penalty term $\lambda_{\mathrm{size}}\sum_i x_i$, giving $\min_x \sum_i (F_{\mathrm{cost}}(t_i)-F_{\mathrm{div}}(t_i)+ \lambda_{\mathrm{size}})x_i$ with $\lambda_{\mathrm{size}}=1$ in this trace; and \blueprintAgent keeps the same QAOA workflow to TCM.
Generally, \ApproachName can distinguish between TCS and TCM requirements in the QUBO coefficients, where TCM includes the size penalty, while TCS does not.



\subsubsection{Code Generation and Repair}

\textbf{\codegenAgent} writes and checks a program for the planned quantum
execution through a within-agent loop, using the tool support (Section~\ref{subsec:tooling}). The program reads the current subproblem from the
execution environment, so code generation happens once for a sample block while
downstream execution can run the same program over all encoded subproblems. In
the loop, the agent may consult coding knowledge, write code, and run static
and compile checks. It may also run the candidate on a sample subproblem in
the sandbox and revise it from tool feedback until the program becomes
executable or the budget is exhausted. Because the same program is then reused
across encoded subproblems, the LLM does not need to regenerate code for every
block. This trace is recorded in the generation record
(Section~\ref{subsubsec:generation_record}), including the generated program
and the tool outcomes used to repair it.

\textbf{\reviewerAgent} handles a cross-agent repair loop before downstream
execution, because a semantically misaligned program can waste execution cost and
produce misleading trace records. It reads the original requirement together
with the structured task description, kernel specification, workflow
blueprint, encoding plan, and candidate program to decide whether execution
should proceed. This review is necessary because executable code can still
violate requirement-level semantics. It is a structured semantic review rather
than another tool-execution loop: static checks and sandbox runs have already
been attempted by \codegenAgent. It targets failures that basic execution
checks may miss, such as stubs, objective mismatch, or replacing the selected
quantum kernel with a simpler routine. If the verdict is \texttt{revise}, it
returns concrete feedback to \codegenAgent; \codegenAgent rewrites the
program, and \reviewerAgent checks the new version within a bounded number of
rounds. This makes review-driven repair observable in the generation record
before downstream execution. If the review budget is exhausted, the workflow
can still proceed with the latest code, and \verificationAgent later records
whether the executed result is supported. \reviewerAgent is therefore critical
as the last requirement-level check before execution is scaled to all encoded
subproblems.

\subsubsection{Output Analysis and Verification}

\textbf{\combinationAgent} takes the subproblem results produced by executing a partitioned problem and combines them into a single value for the original instance. The combination stage applies the rule selected by \encodingAgent. When the rule is a simple numeric aggregation over numeric subproblem outputs, such as sum or mean, the orchestrator uses a deterministic script. When the rule requires semantic interpretation of the subproblem results or cannot be expressed using these fixed aggregators, the orchestrator invokes the LLM combination agent. This hybrid design keeps the LLM responsible for selecting and justifying the combination strategy in the workflow plan, while deterministic code handles straightforward arithmetic whenever possible.

\textbf{\verificationAgent} closes the link between the generated quantum
application and the generation record. Unlike \reviewerAgent, which checks
code before execution, \verificationAgent checks the executed application and
its combined value after execution. It checks whether the code still matches
the declared kernel and, when the returned instance is supported by the
classical reference solver in Section~\ref{subsec:tooling}, compares the
reported value against that reference. We use this classical solver to improve
\ApproachName's result validation, while still aiming to produce executable
quantum applications that can eventually run on quantum hardware. The status is
written to the generation record and attached to the application output.
Cross-checked cases match the classical reference within tolerance. Unverified
cases use an instance schema outside the solver's support. Mismatch cases
disagree with the reference beyond tolerance. Other outcomes record
verification failure: off-spec, when the code falls outside the declared
kernel, or failed, when no terminal verdict is produced. Thus, the workflow
records the evidence supporting the generated application.
\subsection{Supporting Capabilities of \ApproachName}
\label{subsec:supporting_capabilities}
\ApproachName includes supporting capabilities for two recurring needs: one is the task knowledge that helps interpret NL requirements as quantum-suitable formulations, and the other is the coding skill that contributes to the implementation of selected kernels and encodings into executable programs.
For the studied \TCO tasks, a lightweight combination of task knowledge and coding skill is sufficient because the formulation patterns (e.g., QUBO, Ising, and Hamiltonian forms) and the implementation conventions based on Qiskit and Qiskit-Aer remain stable across the requirements in our benchmarks; see Section~\ref{sec: conclusion} for future generalization.

\subsubsection{Task Knowledge}
Task knowledge in \ApproachName includes two aspects: formulation guidance and requirement grounding.
%
\textit{Formulation guidance} helps identify the decision variables, optimization objective, and requirement parameters that must be preserved in the formulation. It supports the construction of QUBO- or Ising-style formulations, qubit estimates, and the subproblem interface passed to code generation.
\textit{Requirement grounding} ties the formulation to the entities and fields that are valid for the targeted SE optimization task. For \TCO, this includes test cases and their available objective fields, such as execution cost or a requested selection size when the requirement provides one. This grounding keeps the formulation close to the supplied requirement, avoids introducing unsupported fields or constraints, and provides the vocabulary that the parse and analysis stages use when extracting decision variables and objective terms from the NL requirement.

\subsubsection{Coding Skill}
Coding skill in \ApproachName includes two aspects: implementation practice and workflow discipline.
\textit{Implementation practice} captures implementation-level conventions for generated quantum applications. The generated code should faithfully implement the upstream kernel and encoding, using QUBO, Ising, or Hamiltonian representations when appropriate. It should use APIs available in the configured Python and Qiskit runtime, avoid stubs, hard-coded instance values, and deprecated library calls, and reuse the low-level support exposed through the building-blocks API manifest in Section~\ref{subsec:tooling} when appropriate.
\textit{Workflow discipline} constrains how code generation proceeds in \ApproachName. Specifically, the code-generation stage is expected to produce a complete program, keep it as the candidate kernel code, and revise the same artifact in response to safety checks, compilation results, and sandbox execution feedback. This discipline makes the workflow produce a checked program that can be reused by execution, combination, and verification, rather than a detached circuit sketch.

\section{Empirical Study}\label{sec: empirical}
\subsection{Research Questions (RQs)}\label{sec: research_questions}
Since we aim to examine whether \ApproachName can turn NL requirements into executable quantum applications, we define the following five RQs:
\begin{enumerate}[label=\textbf{RQ\arabic*}, leftmargin=*]
	\item \textbf{Generation performance}: How effective and efficient is \ApproachName at generating quantum applications that compile, run, and return results?
	
	\item \textbf{Solution quality}: How good are the solutions returned by \ApproachName-generated quantum applications for solving \TCO problems?

	\item \textbf{Pattern investigation}: What workflow patterns can be observed in \ApproachName generation traces?

	\item \textbf{Workflow ablation}: How do components of \ApproachName play a role in quantum application generation?

    \item \textbf{Backbone LLM comparison}: How does \ApproachName perform across different backbone LLMs?
\end{enumerate}

\subsection{Benchmarks}\label{sec: benchmarks}
We evaluate \ApproachName on 10 real-world test-suite benchmarks for TCS and TCM, drawn from prior quantum \TCO studies~\cite{wang2024quantum,trovato2024reformulating}. 
Following the requirement template described in Section~\ref{sec: user_requirements}, we use 20 NL requirements in total, each associated with one benchmark instance and one task specification: TCS or TCM. 
For each requirement, \ApproachName receives \textit{only} the NL prompt and \textit{autonomously} decides, through its prompted agents, how to formulate a problem and generate the corresponding quantum application.
Table~\ref{tab: benchmark} summarizes benchmarks.
The first six were used by Wang et al.~\cite{wang2024quantum}: ABB paint-robot test data (\texttt{paintcontrol}, \texttt{iofrol})~\cite{spieker2018atcsdata}, the Google Shared Dataset of Test Suite Results (\texttt{gsdtsr})~\cite{googleSharedDataset}, and Orona's elevator-dispatcher test scenarios (\texttt{elevator\_o2}, \texttt{elevator\_o3})~\cite{valle2023delta}. We also include \texttt{sampled\_paintcontrol} as a reduced \texttt{paintcontrol} variant. The remaining four benchmarks, \texttt{flex}, \texttt{grep}, \texttt{gzip}, and \texttt{sed}, are GNU subject programs from the SIR repository~\cite{do2005supporting}, used by Trovato et al.~\cite{trovato2024reformulating}.
\begin{table}[htbp]
    \footnotesize
    \renewcommand{\arraystretch}{0.9}
    \caption{Benchmarks used in the empirical study}
    \label{tab: benchmark}
    \begin{center}
        \resizebox{0.88\columnwidth}{!}{
            \begin{tabular}{c|l|l|c|cc}
    \toprule[1pt]
    \multicolumn{1}{c|}{\textbf{BID}} &
    \multicolumn{1}{c|}{\textbf{Benchmark name}} &
    \multicolumn{1}{c|}{\textbf{Objective Attributes}} &
    \multicolumn{1}{c|}{\textbf{\# Tests}} &
    \multicolumn{1}{c}{\textbf{RID$_{\mathrm{TCS}}$}} &
    \multicolumn{1}{c}{\textbf{RID$_{\mathrm{TCM}}$}} \\
    \cmidrule(lr){1-1} \cmidrule(lr){2-2} \cmidrule(lr){3-3} \cmidrule(lr){4-4} \cmidrule(lr){5-5} \cmidrule(lr){6-6}
    0 
    & \texttt{elevator\_o2}
    & cost, div.
    & 1,925
    & 0
    & 10
    \\
    1
    & \texttt{elevator\_o3}
    & cost, pass., dist.
    & 1,925
    & 1
    & 11
    \\
    2
    & \texttt{gsdtsr}
    & cost, size, fail
    & 5,555
    & 2
    & 12
    \\
    3
    & \texttt{iofrol}
    & cost, size, fail
    & 1,941
    & 3
    & 13
    \\
    4
    & \texttt{paintcontrol}
    & cost, size, fail
    & 90
    & 4
    & 14
    \\
    5
    & \texttt{sampled\_paintcontrol}
    & cost, size, fail
    & 46
    & 5
    & 15
    \\
    6
    & \texttt{flex}
    & cov., cost, fault
    & 567
    & 6
    & 16
    \\
    7
    & \texttt{grep}
    & cov., cost, fault
    & 806
    & 7
    & 17
    \\
    8
    & \texttt{gzip}
    & cov., cost, fault
    & 214
    & 8
    & 18
    \\
    9
    & \texttt{sed}
    & cov., cost, fault
    & 360
    & 9
    & 19
    \\
    \bottomrule[1pt]
\end{tabular}
        }
        {\justify\footnotesize
            BID and RID are the benchmark and requirement IDs, respectively.
            For Objectives,
            cost = test-execution cost; div.\ = diversity of passenger-traffic input scenarios; pass.\ = passenger count; dist.\ = travel distance; fail = failure rate; fault = past-fault coverage.
        \par}
    \end{center}
\end{table}

\subsection{Evaluation Metrics}\label{sec: metrics}
Generation records offer specific and reliable metadata to calculate metrics in terms of the following three dimensions.

\textbf{Generation cost.}
We measure generation cost with time and token consumption: the former is the duration of a workflow run, and the latter is computed from all tokens consumed during generation. We report both metrics for each requirement and average valid records across workflow runs.

\textbf{Generation effectiveness.}
We measure \ApproachName's effectiveness using three pass@$k$ metrics for executable application artifacts: compilation (Cpl.), execution (Exe.), and final combination (Cbi.), which respectively check whether \codegenAgent produces Python code that passes compilation; a generated application runs successfully in the sandbox; and the workflow returns a final numeric output after postprocessing, including aggregation of decomposed subproblem outputs when decomposition is used. 
Furthermore, RQ5 reports the end-to-end correctness of each phase included in \ApproachName.
Identification of success is based on clear records in the logs.
For $n$ repeated runs with $c$ successful runs, we compute pass@$k$ as $1-\binom{n-c}{k}/\binom{n}{k}$~\cite{chen2021evaluating}.
Owing to the per-stage record, we can further zoom into the success proportion of 8 individual agents introduced by \ApproachName, enabling a more fine-grained analysis of application generation.

\textbf{Solution quality.}
Generation effectiveness and solution quality measure different properties: a workflow may produce an executable quantum application, yet still return a poor solution with respect to the optimization objective. We therefore use QUBO-value reduction ratio $QRR$ to measure solution quality.
We define $QRR$ on QUBO values because the studied \TCO tasks are binary optimization problems, and QUBO provides a common scalar objective for comparing generated solutions across requirements and against the baseline under the same encoding. This choice also covers equivalent Ising formulations, which can be transformed to QUBO form without changing the underlying optimization target.

Specifically, let $qubo_{\text{Q}}$ denote the minimum valid QUBO value $\min_{i=1}^{n'} \mathcal{O}_{\text{Q}}^{i}(\mathbf{s}^{i}_{\text{Q}})$ jointly determined by $n' (n' \le n)$ valid solutions $\{\mathbf{s}_{\text{Q}}^{i}\}_{i=1}^{n'}$ and QUBO formulations $\{\mathcal{O}^{i}_{\text{Q}}\}_{i=1}^{n'}$ from $n$ repeated runs of \ApproachName;
Then, let $qubo_{\text{C}}$ denote a minimum valid QUBO value $\min_{i=1}^{n'} \mathcal{O}_{\text{Q}}^{i}(\mathbf{s}_{\text{C}})$ induced by a classical algorithm, where the classical counterpart returns one solution $\mathbf{s}_{\text{C}}$ and we reuse $\{\mathcal{O}^{i}_{\text{Q}}\}_{i=1}^{n'}$ to make the output type uniform for fair comparison between quantum and classical solutions.
Finally, for each problem, we compute $QRR := (qubo_{\text{C}} - qubo_{\text{Q}}) / |qubo_{\text{C}}|$.
%
For this classical reference, we use the GA baseline from recent QAOA-based \TCO work~\cite{wang2024quantum}, reuse its configuration, and evaluate its solutions with the same QUBO objective as \ApproachName. A positive, zero, or negative $QRR$ means that \ApproachName is better than, matches, or is worse than this reference.
Unlike \verificationAgent's built-in solver over sampled solutions, $QRR$ uses the GA baseline, which solves the full optimization problem outside the \ApproachName workflow.

\subsection{Experimental Settings}\label{sec: settings}
\textbf{Backbone LLMs.}
Our study uses three backbone LLMs popular for coding tasks:
\texttt{claude-opus-4-7} (Claude)~\cite{anthropicClaudeModels2026},
\texttt{DeepSeek-V4-Flash} (DeepSeek)~\cite{xu2026deepseek}, and
\texttt{llama4-maverick-17b} (Llama)~\cite{metaLlama4ModelCard2025}, where RQ1--RQ4 are fixed with Claude while RQ5 compares among the three. These models are selected to cover different provider ecosystems and coding-oriented options for agent workflows. 
For open-weight LLMs, we do not override decoding controls and simply use default settings.

\textbf{Baselines.}
We introduce four ablation variants/baselines in RQ4. \AblationSingle is designed to assess the necessity of a multi-agent workflow, and it adopts the same workflow as \ApproachName but merely uses a single agent prompted with all agent roles.
Supporting capabilities play a critical role in offering task-specific preliminaries, such that \AblationWithoutSkill without codegen skills and \AblationWithoutKnowledge without task knowledge are considered to assess their roles in \ApproachName. 
Since \reviewerAgent is a critical component to improve the codegen quality through iterative feedback, baseline \AblationWithoutReview is designed with this agent removed.

\textbf{Problem selection.}
RQ1--RQ3 use Claude only and cover all 20 requirements.
To keep the cost of running multiple LLMs and ablation baselines manageable while preserving problem complexity, RQ4 and RQ5 sample the top 10 requirements ranked by the number of test cases, including 5 TCS and 5 TCM requirements.

\textbf{Execution settings.}
\textit{Repetition:} For each requirement, we run the same staged pipeline 3 times ($n=3$), using a batch runner with four concurrent runs and a 1,500-second timeout per run.
\textit{Budget:} We fix the execution environment and resource budgets, while the generated program determines the simulation routine and sampling parameters when applicable.
We set a 24-qubit budget to keep runs simulator-feasible and trigger decomposition for larger requirements, with a 2,048 MB memory limit and a 60-second sandbox timeout to bound invalid or overly expensive programs. These settings are configurable, and \ApproachName can be adapted to other quantum backends and resource budgets.
\textit{Environment:}
\ApproachName is implemented in Python 3.11.
Generated applications target Qiskit 2.x and run in a local, noise-free simulator rather than on quantum hardware.

\subsection{Experimental Results}
\subsubsection{RQ1: Generation Evaluation}\label{sec: rq1}

RQ1 evaluates whether \ApproachName can produce executable quantum applications
from the 20 NL requirements.
We report requirement-level averages for cost and pass@1 for
compilation, execution, and final combination. 
We use pass@1 as it is the strictest single-run success measure from the planned $n$ runs, whereas larger $k$ values credit success across multiple attempts.

\begin{table}[htbp]
    \footnotesize
    \renewcommand{\arraystretch}{1}
    \caption{Generation cost and effectiveness by TCO problem.}\label{tab: rq1_performance}
    \begin{center}
        \resizebox{0.58\columnwidth}{!}{
            \begin{tabular}{c|cc|ccc}
    \toprule[1pt]
    \textbf{Case} & \textbf{Time (s)} & \textbf{Token (M)} & \textbf{Cpl.} & \textbf{Exe.} & \textbf{Cbi.} \\
    \cmidrule(lr){1-1} \cmidrule(lr){2-2} \cmidrule(lr){3-3} \cmidrule(lr){4-4} \cmidrule(lr){5-5} \cmidrule(lr){6-6}
    BID-0 & 266.3 & 1.70 & 1.000 & 0.667 & 0.667 \\
    BID-1 & 300.3 & 1.99 & 1.000 & 1.000 & 1.000 \\
    BID-2 & 463.6 & 2.82 & 1.000 & 1.000 & 1.000 \\
    BID-3 & 300.3 & 1.98 & 1.000 & 1.000 & 1.000 \\
    BID-4 & 190.0 & 1.64 & 1.000 & 1.000 & 1.000 \\
    BID-5 & 179.5 & 1.66 & 1.000 & 1.000 & 1.000 \\
    BID-6 & 214.6 & 1.77 & 1.000 & 1.000 & 1.000 \\
    BID-7 & 227.0 & 1.82 & 1.000 & 1.000 & 1.000 \\
    BID-8 & 237.5 & 1.73 & 1.000 & 1.000 & 1.000 \\
    BID-9 & 222.2 & 1.74 & 1.000 & 1.000 & 1.000 \\
    \cmidrule(lr){1-1} \cmidrule(lr){2-2} \cmidrule(lr){3-3} \cmidrule(lr){4-4} \cmidrule(lr){5-5} \cmidrule(lr){6-6}
    TCS & 268.1 & 1.87 & 1.000 & 0.933 & 0.933 \\
    TCM & 252.2 & 1.91 & 1.000 & 1.000 & 1.000 \\
    \cmidrule(lr){1-1} \cmidrule(lr){2-2} \cmidrule(lr){3-3} \cmidrule(lr){4-4} \cmidrule(lr){5-5} \cmidrule(lr){6-6}
    All & 260.1 & 1.89 & 1.000 & 0.967 & 0.967 \\
    \bottomrule[1pt]
\end{tabular}
        }
        {\justify\footnotesize
            Each benchmark case (BID-0 to BID-9) averages its paired TCS and TCM
            requirements. Time and Token report requirement-level averages.
            Compilation, execution, and final combination report pass@1 for the
            corresponding generation-effectiveness metrics.
        \par}
    \end{center}
\end{table}

As shown in Table~\ref{tab: rq1_performance}, \ApproachName can successfully generate executable applications for almost all requirements, where failed execution and compilation occur only in \texttt{elevator\_o2} (BID-0) for the corresponding TCS problem. The remaining failures occur after compilation, indicating that execution-time behavior and output integration are the main sources of failures.
Overall pass@1 reaches 1.000 for compilation and 0.967 for both execution and final combination, with average generation costs of 260.1~seconds and 1.89M tokens per requirement. These results show that most generated applications are syntactically valid and can be executed to produce final outputs.
Similar costs across TCS and TCM further suggest that both problem variants place comparable demands on the workflow.

\begin{figure}[htbp]
    \centering
    \includegraphics[width=0.68\columnwidth]{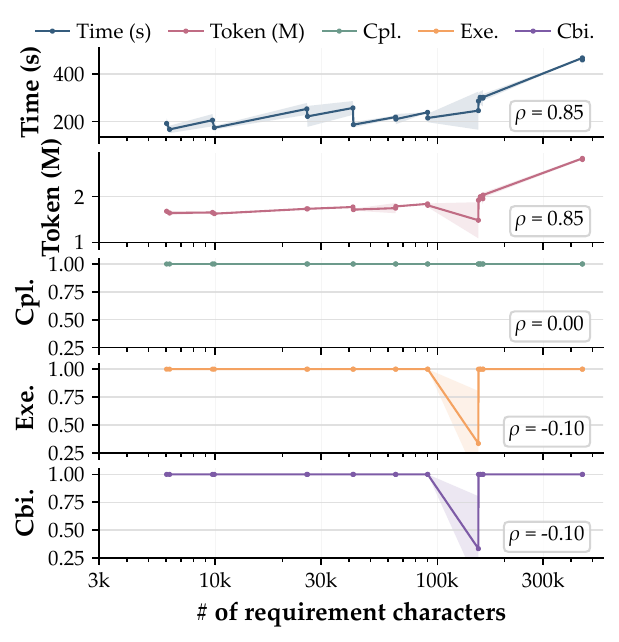}
    \caption{Relation between lengths of 20 requirements and RQ1 metrics.}\label{fig: rq1_requirement_correlation}
\end{figure}

Figure~\ref{fig: rq1_requirement_correlation} relates requirement length, measured by the number of characters in the NL requirement, to the RQ1 metrics using Spearman's $\rho$. Requirement length is strongly correlated with generation cost ($\rho=0.85$ for both time and token consumption), but shows little association with generation-effectiveness metrics: compilation is unchanged ($\rho=0.00$), and execution and combination have only weak negative correlations ($\rho=-0.10$). Thus, longer requirements mainly increase processing cost rather than systematically reducing generation success. The observed success drops (related to BID-0 in Table~\ref{tab: rq1_performance}) are more likely due to a small number of harder TCS cases than to requirement length itself.

\AnswerToRQ{1}{On the studied 20 \TCO requirements, \ApproachName achieves pass@1 of 1.000 for compilation and 0.967 for both execution and final combination, with average generation cost of 260.1 seconds and 1.89M tokens per requirement.}

\subsubsection{RQ2: Solution Quality}\label{sec: rq2}


RQ2 examines whether the quantum applications successfully generated by \ApproachName can return competitive optimization results, using the $QRR$ metric (Section~\ref{sec: metrics}) to compare against GA.

\begin{figure}[htbp]
    \centering
    \includegraphics[width=0.62\columnwidth]{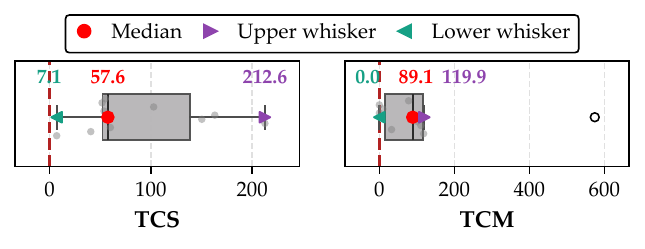}
    \caption{Distribution of $QRR$ across successful generations of 10 TCS and 10 TCM requirements}\label{fig: rq2_boxplot}
\end{figure}

Figure~\ref{fig: rq2_boxplot} shows positive $QRR$ distributions for both TCS and TCM. 
TCM has a higher median $QRR$ (89.1) and a tighter spread, suggesting more consistent improvements. TCS has a lower median (57.6) but a larger upper whisker (212.6), indicating higher variability and stronger best-case reductions.
The lower whiskers show that the weakest observed cases are still near or above the GA reference: for TCS, $QRR$ remains positive at 7.1, while for TCM the lower end reaches 0.0. 
This finding is significant because \ApproachName can automatically and effectively generate useful solutions for TCO problems. Compared with prior quantum approaches that rely on manually designed problem-solving pipelines~\cite{wang2024quantum,trovato2024reformulating,wang2024test}, \ApproachName achieves performance comparable to, and in most cases better than, classical GA baselines.
Besides, \ApproachName's high-quality solutions may be partially attributed to \verificationAgent's small-scale comparison between the LLM-generated solution and the built-in classical solver, which can reduce the risk of returning poor-quality solutions to users.

\AnswerToRQ{2}{Among successful generations, \ApproachName never underperforms the GA baseline under $QRR$: the worst observed cases are $QRR=7.1$ for TCS and $QRR=0.0$ for TCM, while the medians are $QRR=57.6$ for TCS and $QRR=89.1$ for TCM.}

\subsubsection{RQ3: Pattern Investigation}\label{sec: rq3}

RQ3 analyzes workflows obtained in the 58 successful generations, out of 60 total runs (20 requirements $\times$ 3 repeats).
We use only successful generations because pattern extraction requires complete traces from \analysisAgent through \verificationAgent.
The statistics are based on runs rather than requirements, because for one requirement, \ApproachName may choose different patterns in different runs, such as RID-7 that plans workflows with 4, 5, and 7 nodes in three successful runs.

\begin{figure}[htbp]
    \centering
    \includegraphics[width=0.72\columnwidth]{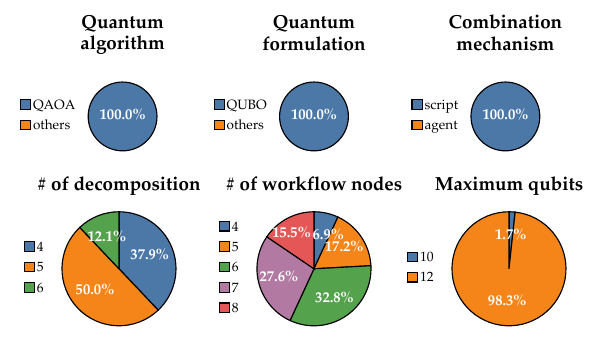}
    \caption{Pie charts for six patterns observed in 58 successful runs of \ApproachName}\label{fig: rq3_pattern}
\end{figure}

Figure~\ref{fig: rq3_pattern} shows that the successful runs share a stable prompt-conditioned quantum core: all 58 use a QUBO formulation and QAOA. 
This pattern is \textit{not} from predefined workflow settings or fixed model configurations.
Rather, in each run, agents autonomously choose the formulation and algorithm, and all 58 successful runs happen to converge on QUBO and QAOA for our target problems, even though we prompt \analysisAgent with exemplary algorithms previously used as solutions to SE optimization problems~\cite{zhang2026quantum}, like variational quantum eigensolver and Grover search, while no specified encoding candidates are offered for \encodingAgent.
The fact that all successful runs use QUBO also does not mean that Ising is disabled or unavailable, since the two are interchangeable binary representations and agents can convert between them with \texttt{read\_blocks\_api} in Table~\ref{tab:tooling}.
Furthermore, we observe that agent-generated QUBO formulations differ from those in prior work~\cite{wang2024quantum}, where agent-generated QUBO formulations for TCM introduce penalty terms to confine the number of selected test cases as discussed in Section~\ref{sec: requirements and encoding}, while the prior study models this aspect as an objective to be minimized.
The ``Combination mechanism'' pie chart shows that \ApproachName's \combinationAgent autonomously chooses between script-based and agent-based final aggregation, and selects the script-based option in all 58 successful runs.

The main variation appears in the structure of a generated application workflow. In Figure~\ref{fig: rq3_pattern}, the pie chart of ``\# of decompositions'' shows that successful runs split the problem into 4, 5, or 6 subproblems, and most (37.9\%+50.0\%=87.9\%) of the successful runs adopt decompositions below 6.
The pie chart of ``\# of workflow nodes'', where a node denotes one step in the agent-generated application workflow, further characterizes the resulting orchestration structure. Workflows with 6 and 7 nodes are the most common and together account for about 60\% of successful runs, with 8-node workflows also appearing in a non-trivial fraction. This indicates that successful generations usually require a multi-step workflow with classical preparation, quantum execution, postprocessing, and combination, rather than a minimal single quantum call.
This suggests that effective quantum-application generation in \ApproachName depends not only on selecting a suitable quantum kernel, but also on orchestrating several supporting steps around it.
The pie chart of ``Maximum qubits'' shows that almost all runs keep each circuit at 12 qubits, despite the 24-qubit budget in Section~\ref{sec: settings}.
Taken together, these results indicate that \encodingAgent balances two competing concerns. It preserves simulation feasibility by decomposing large \TCO instances into moderately sized quantum executions, rather than driving individual circuits close to the resource limit. At the same time, by keeping most circuits at 12 qubits instead of using even smaller circuits, it limits the number of subproblems and thus avoids excessive serial subproblem-solving overhead.

\AnswerToRQ{3}{\ApproachName's advantage is that its agents autonomously construct resource-aware workflows rather than relying on fixed templates: across successful runs, they consistently converge on executable QUBO+QAOA solutions, decompose tasks into 4 to 6 subproblems, keep most circuits at 12 qubits for feasibility, and assemble them into 6- or 7-node workflows. This shows that \ApproachName can systematically turn a requirement into an executable quantum-application workflow under practical resource constraints.}

\subsubsection{RQ4: Workflow Ablation}\label{sec: rq4}

\begin{figure}[htbp]
    \centering
    \includegraphics[width=0.82\columnwidth]{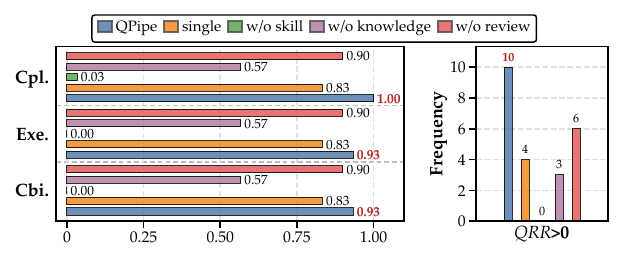}
    \caption{Ablation results: the left part reports pass@1 for \textit{Cpl.}, \textit{Exe.}, and \textit{Cbi.}, while the right part evaluates solution quality.}\label{fig: rq4_ablation}
\end{figure}

As shown in the left part of Figure~\ref{fig: rq4_ablation}, \ApproachName achieves the best generation effectiveness among the workflows instantiated by \ApproachName and its ablation baselines: its \textit{Cpl.} reaches 1.00, while \textit{Exe.} and \textit{Cbi.} remain close to 1.00. In contrast, all ablation workflows show lower effectiveness. The largest degradation appears in \AblationWithoutSkill, which almost never passes \textit{Cpl.} and never reaches \textit{Exe.} or \textit{Cbi.}, showing that \codegenAgent skills are essential for executable quantum-application generation. \AblationWithoutKnowledge also drops clearly on all three metrics, while \AblationSingle and \AblationWithoutReview remain closer to \ApproachName but still perform worse. This suggests that task knowledge, specialized agents, and review feedback each contribute to workflow stability.

The right part of Figure~\ref{fig: rq4_ablation} shows that \ApproachName achieves the strongest solution-quality performance under $QRR$.
\ApproachName returns valid solutions for all 10 requirements, all with positive $QRR$.
In comparison, fewer requirements under \AblationSingle and \AblationWithoutReview achieve positive $QRR$; \AblationWithoutKnowledge further reduces this number, and \AblationWithoutSkill returns no solution better than the classical reference. Thus, \ApproachName not only generates more executable applications, but also more reliably produces high-quality solutions.

\AnswerToRQ{4}{\ApproachName outperforms all ablations in both generation effectiveness and positive $QRR$ coverage. Removing codegen skills causes the largest drop, while removing task knowledge, review feedback, or the multi-agent decomposition also degrades performance.}

\subsubsection{RQ5: Backbone LLM Comparison}\label{sec: rq5}

RQ5 evaluates how \ApproachName performs with three backbone LLMs.
Each model is run three times per requirement under the protocol in Section~\ref{sec: settings}. We focus on generation effectiveness rather than solution quality because solution quality is only observable for successfully generated applications; when a backbone fails on many requirements, its quality results would cover only a small and potentially biased subset.

\begin{table}[htbp]
    \footnotesize
    \renewcommand{\arraystretch}{0.98}
    \caption{Backbone LLM generation effectiveness on the selected 10 requirements.}\label{tab: rq5_cross_model}
    \begin{center}
        \resizebox{0.98\columnwidth}{!}{
            \begin{tabular}{c|ccc|ccc|ccc|cccccccc}
    \toprule[1pt]
    \textbf{LLM} & \multicolumn{3}{c|}{\textbf{pass@1}} & \multicolumn{3}{c|}{\textbf{pass@2}} & \multicolumn{3}{c|}{\textbf{pass@3}} & \multicolumn{8}{c}{\textbf{Agent-level success}} \\
    \cmidrule(lr){2-4} \cmidrule(lr){5-7} \cmidrule(lr){8-10} \cmidrule(lr){11-18}
    & \textbf{Cpl.} & \textbf{Exe.} & \textbf{Cbi.} & \textbf{Cpl.} & \textbf{Exe.} & \textbf{Cbi.} & \textbf{Cpl.} & \textbf{Exe.} & \textbf{Cbi.} & \textbf{Par.} & \textbf{Ana.} & \textbf{Blue.} & \textbf{Enc.} & \textbf{Code.} & \textbf{Rev.} & \textbf{Cbi.} & \textbf{Ver.} \\
    \cmidrule(lr){1-1} \cmidrule(lr){2-2} \cmidrule(lr){3-3} \cmidrule(lr){4-4} \cmidrule(lr){5-5} \cmidrule(lr){6-6} \cmidrule(lr){7-7} \cmidrule(lr){8-8} \cmidrule(lr){9-9} \cmidrule(lr){10-10} \cmidrule(lr){11-11} \cmidrule(lr){12-12} \cmidrule(lr){13-13} \cmidrule(lr){14-14} \cmidrule(lr){15-15} \cmidrule(lr){16-16} \cmidrule(lr){17-17} \cmidrule(lr){18-18}
    Claude & 1.000 & 0.933 & 0.933 & 1.000 & 0.967 & 0.967 & 1.000 & 1.000 & 1.000 & 30/30 & 30/30 & 30/30 & 30/30 & 30/30 & 28/30 & 28/30 & 28/30 \\
    DeepSeek & 1.000 & 0.500 & 0.500 & 1.000 & 0.667 & 0.667 & 1.000 & 0.800 & 0.800 & 30/30 & 30/30 & 30/30 & 30/30 & 30/30 & 15/30 & 15/30 & 14/30 \\
    Llama & 0.000 & 0.000 & 0.000 & 0.000 & 0.000 & 0.000 & 0.000 & 0.000 & 0.000 & 30/30 & 30/30 & 30/30 & 30/30 & 0/30 & 0/30 & 0/30 & 0/30 \\
    \bottomrule[1pt]
\end{tabular}
        }
    \end{center}
    {\justify\footnotesize
        ``Par.'', ``Ana.'', ``Blue.'', ``Enc.'', ``Code.'', ``Rev.'', ``Cbi.'', and ``Ver.'' under ``Agent-level success'' correspond to the parse, analysis, blueprint, encoding, codegen, review, combination, and verification phases, respectively. An entry ``$x$/30" indicates $x$ successful runs out of 30 executions (10 requirements $\times$ 3 repeats)
    \par}
\end{table}

As shown in Table~\ref{tab: rq5_cross_model}, \textbf{Claude} achieves the strongest results, with pass@1 of 1.000 for compilation and 0.933 for both execution and final combination, and pass@3 of 1.000 across all stages. Its agent-level results are also stable: all runs pass parsing, analysis, blueprinting, encoding, and code generation, with only two failures observed in review, combination, and verification. This suggests that the staged workflow is effective when the backbone model consistently preserves task constraints, intermediate decisions, and artifact dependencies across agents.
\textbf{DeepSeek} also supports substantial progress through the workflow. It matches Claude in the early agents and compilation, achieving full success through code generation and 1.000 pass@3 for compilation. Review and combination succeed in about half of the runs, with verification slightly lower, leading to lower execution and final-combination pass@$k$. This shows that repeated attempts can recover compilation success, while later runtime-facing stages remain more sensitive to backbone behavior.
\textbf{Llama} completes parsing, analysis, blueprinting, and encoding in all runs, showing that the early workflow stages can still produce structured intermediate artifacts. In these runs, however, applications do not pass the code-generation agent, so downstream execution and final-combination pass@$k$ remain zero.

These results suggest that early agents can often produce stable intermediate artifacts across backbone models, while larger differences appear in code generation and in the later review, combination, and verification stages. This indicates that future improvement should focus on tuning these runtime-facing agents for specific models or model families. We expect that better adapted review, combination, and verification phases could improve end-to-end performance. Other factors, such as prompt design, context length, tool feedback, retry strategy, and model-specific coding behavior, may also affect performance and deserve further investigation.

\AnswerToRQ{5}{Claude achieves the strongest end-to-end performance, DeepSeek usually reaches compilation but drops at later stages, and Llama fails at code generation. Backbone differences therefore concentrate in code generation and the later review, combination, and verification stages rather than in early planning stages.}

\subsection{Threats to Validity}\label{sec: threats}

\textbf{Internal validity.}
Results depend on the current implementation, prompts, agent settings, and execution environment of \ApproachName.
We cannot guarantee bug-free implementation, but the tool and benchmark artifacts are available~\cite{qpipe_tool_link,qpipe_empirical_link} for inspection and reproduction.
Remote LLM services may vary with service load, model updates, or network latency; we mitigate this by using the same protocol and machine-side configuration throughout.

\textbf{Construct validity.}
Our metrics capture whether \ApproachName generates executable quantum applications and at what cost, but not every aspect of application quality.
For solution quality, $QRR$ compares returned QUBO values with the GA baseline for the same generated formulation; it is a relative formulation-level measure, not an absolute task-specific quality measure.
We evaluate agent reasoning over given NL requirements, not the quality of the requirement construction process itself.
Data leakage is another potential threat, although the sandbox  can block access to online artifacts during generation. Besides, benchmark instances or related artifacts may have appeared in an LLM's training data, but the task is not to reproduce a fixed formulation or a known algorithm, where the generated application and its selection depend on reasoning over the supplied requirement and instance. Because each run produces a new application, prior exposure to an artifact does not directly provide the generated solution.

\textbf{External validity.}
The evaluation covers 10 real-world test-suite benchmarks, 20 associated NL requirements, and two \TCO variants from prior quantum \TCO studies~\cite{wang2024quantum,trovato2024reformulating}. In these TCS and TCM problems, coverage and selection size are optimization objectives rather than hard constraints. The resulting requirements therefore provide substantial evidence about objective-driven application generation, but limited evidence about complex constraint handling; this reflects the benchmark scope, rather than an assumption in \ApproachName. Because quantum formulations often encode constraints with penalty terms, direct classical--quantum comparison on the same QUBO can also be unfair. Constraint-intensive SE problems should therefore be included in future evaluations.

In addition, the results may not generalize to other SE optimization tasks, benchmark families, quantum backends, or requirement styles. The requirements do not prescribe a target backend, and the agents consequently select ideal simulators in the current setup. This choice means that the present evaluation does not measure how the agents would select or configure noisy devices, account for device constraints, or perform hardware-specific compilation when such requirements are provided. The backbone and baseline comparisons are also limited to the studied LLMs, ablations, and GA reference. 
One reason could be no prior study offering accessible LLM-based agentic workflows to generated quantum applications, and GA as the popular classical counterpart used in the literature that explored quantum optimization for test optimization~\cite{wang2024quantum}.

\textbf{Conclusion validity.}
LLM-based generation is stochastic, and the evaluation budget limits repetitions, backbones, and ablation settings.
We repeat each run three times and use standard metrics such as pass@1, but the conclusions are evidence for the studied setting, not final claims about all models, benchmarks, baselines, or quantum-application generation workflows.

\section{Related Work}\label{sec: related}
\textit{SE for QC} studies how SE methods should be adapted to develop quantum software systems~\cite{zhao2020quantum,murillo2025quantum,ali2022software}, including software testing~\cite{li2026methodological} and maintenance~\cite{chen2023smelly}.
\textit{QC for SE}, instead, applies QC to address SE problems~\cite{zhao2025quantum,pezze20252030}, especially through quantum optimization~\cite{zhang2026quantum,zhang2025empirical}, such as QA for test-case minimization~\cite{wang2024test} and QAOA for \TCO~\cite{wang2024quantum}.
Other studies explore quantum AI for software engineering (QAI for SE)~\cite{wang2025quantum,de2022qai4ase}, including quantum neural networks~\cite{wang2026quantum}. 
\ApproachName follows this QC-for-SE research line, but focuses on turning NL requirements into executable quantum applications.

\textit{Quantum workflows} capture pipelines that include preprocessing, encoding, backend selection, execution, and postprocessing~\cite{weder2020quantme}.
Weder et al.~\cite{weder2020quantme} introduced \textit{QuantME} as a workflow modeling language for manually specifying preconfigured quantum circuits, which was later applied to Business Process Model and Notation (BPMN) and hence forms the Quantum4BPMN approach. 
Building on QuantME, Beisel et al.~\cite{beisel2025pattern} proposed an approach for generating quantum workflows based on selected quantum computing patterns, including algorithmic patterns such as QAOA and behavioral patterns such as pre-deployed execution. 
Yue and Zhang~\cite{yue2026qready} recently proposed \textit{Q-READY}, a model-based systems engineering/model-driven engineering (MBSE/MDE)-oriented conceptual framework for designing hybrid quantum-classical applications and assessing their feasibility before costly implementation. 
Compared with these works, \ApproachName starts directly from requirements and fully automates the generation, compilation, and execution of quantum workflows, by leveraging a carefully designed feedback-enabled multi-agent architecture.

\textit{LLM-based quantum code generation} has primarily focused on benchmarking, including efforts such as Qiskit HumanEval~\cite{vishwakarma2024qiskit}, QuanBench~\cite{guo2025quanbench}, and QHackBench~\cite{basit2025qhackbench}.
Beyond benchmarking, LLMs have also been used for isolated synthesis tasks, such as QAOA circuit generation~\cite{tyagin2025qaoa} and QUBO transformation from NL requirements~\cite{zhang2025llm}.
While these studies focus on generating correct quantum code, algorithms, or formulations, \ApproachName targets an end-to-end, application-level workflow construction process from SE optimization requirements to executable quantum applications.

\textit{Agents for quantum program development} are emerging, which in general decompose tasks, invoke tools, and use execution feedback to revise intermediate results.
Recent work has explored verifier-in-the-loop quantum program synthesis~\cite{yu2026vista}, agentic reinforcement learning for quantum assembly generation~\cite{yu2025quasar}, multi-agent OpenQASM programming~\cite{fu2025qagent}, multi-agent optimization and error-correction refinement for generated quantum programs~\cite{campbell2025enhancing}, and LLM-based quantum-software debugging~\cite{pham2026qbuglm}.
These efforts demonstrate the value of agents for quantum programming, refinement, and debugging.
\ApproachName instead targets a process from NL requirements to executable quantum applications, rather than focusing on a specific programming language, task, or scope.


\section{Conclusion and Future Work}\label{sec: conclusion}
This paper presented \ApproachName, a multi-agent large language model (LLM) architecture for translating natural language (NL) software engineering (SE) optimization requirements into executable quantum applications, and evaluated it on test case optimization. Overall, \ApproachName reliably produced executable applications for the studied requirements, often returned solutions that match or improve on the genetic-algorithm baseline, and benefited from staged review and verification rather than direct code generation alone. Although the workflow is costly, it can reduce manual effort that would otherwise require SE, quantum, optimization, and programming expertise.

Beyond TCO, future work will adapt \ApproachName's reusable agent architecture and task-specific knowledge to other SE optimization problems, and further study cost, robustness, and execution on real quantum hardware.

\section*{Acknowledgment}\label{sec: acknowledgment}
Apart from serving as a subject in the empirical study, generative AI tools have been used to polish the texts in this paper and to generate scripts for data visualization. All the generated content has been carefully reviewed and edited by the authors.

\bibliographystyle{plainnat}
\bibliography{components/ref}

\end{document}